\documentclass{article}

\usepackage{PRIMEarxiv}
\usepackage{natbib}
\usepackage[utf8]{inputenc} 
\usepackage[T1]{fontenc}    
\usepackage{url}            
\usepackage{booktabs}       
\usepackage{amsfonts}       
\usepackage{nicefrac}       
\usepackage{microtype}      
\usepackage{lipsum}
\usepackage{fancyhdr}       
\usepackage{graphicx}       
\usepackage{amsmath}
\usepackage{booktabs}
\usepackage{caption}
\usepackage{graphicx}
\usepackage{siunitx}
\usepackage{setspace}
\usepackage{subcaption}
\usepackage[table]{xcolor}
\usepackage{rotating}
\usepackage{xcolor}
\usepackage{footnote}
\usepackage{longtable}
\usepackage{hyperref} 
\graphicspath{{media/}}     

\title{The Impact of Geopolitical Risks on Bitcoin Volume Growth: Evidence from a Panel Data Analysis
}

\author{
       Ivan Sergio\thanks{Hamburg Institute of International Economics (HWWI), Oberhafenstrasse 1 20097 Hamburg, Germany, \texttt{sergio@hwwi.org}.} \thanks{Department of Economics, Leibniz-Fachhochschule, Expo Plaza 11, Hannover, 30539, Lower Saxony, Germany, \texttt{ivan.sergio@leibniz-fh.de}.}
       \and
       Danilo Petti\thanks{Corresponding author.}\thanks{School of Mathematics, Statistics and Actuarial Science, University of Essex, Wivenhoe Park, Colchester \texttt{d.petti@essex.ac.uk}.}
}

\begin{document}
\maketitle

\begin{abstract}
This paper investigates the relationship between geopolitical risks (GPR) and the growth rate of Bitcoin (BTC) volume. Our analysis utilizes dynamic panel data from 33 individual countries and the European economic region. Empirical results demonstrate that GPR has a significant positive impact on BTC volume growth, particularly in developing countries. Our results are confirmed by several robustness checks, like Lagged IV, and volatility check among others. Our study offers a new perspective on BTC, as the novelty of the data used helps us understand the dynamics of BTC volume.
\end{abstract}


\section{Introduction}
 \label{ch:introduction}

The meteoric rise of Bitcoin (BTC) as a financial asset has revolutionized the global financial market, prompting extensive academic interest in its integration within the global economy. In 2009, \textit{Bitcoin: A Peer-to-Peer Electronic Cash System} is published under the pseudonym Satoshi Nakamoto \citep{nakamoto2009bitcoin}, marking the beginning of BTC as a digital asset in the world.

To date BTC's inflation rate is about 0.9\%, which is lower than gold’s current rate \citep{WorldGoldCouncil2019}. Literature suggests that this low inflation rate, along with other factors, may elevate Bitcoin to a safe haven asset status, gradually shedding its speculative nature \citep{jrfm17040134, risks9090154, YOUSAF2023103934}.

Bitcoin has previously provided protection against banking crises in some countries \citep{BOURI2017192, BOURI2020156, LUTHER201750}. For example, the BTC price surged during the last European debt crisis (2010–2013) and also during the Cypriot banking crisis (2012–2013) \citep{LUTHER201750}, demonstrating the high capacity of some cryptocurrencies to represent an alternative against political and sovereign risks. The same happened during the minor banking crisis in March 2023, where there were fears of international contagion \citep{jrfm17040134, GALATI2024103001}.

In particular, in an environment of growing geopolitical instability, Bitcoin attract the attention of investors and financial analysts due to its potential predictive power on returns and volatility especially in bearish and bullish market phases \citep{balcilar2017can, sapuric2022relationship, conlon2024bitcoin}, its role in investment decisions as a safe-haven asset \citep{al2020geopolitical, nouir2023economic}, and evidence that shadow market participants shift from cash to cryptocurrency for anonymous transactions, as observed in the Malaysian market \citep{Marmora2021, goel2024cryptocurrency, shahzad2019bitcoin}. These phenomena underscore Bitcoin's potential not only as a financial asset but also as a barometer of market sentiment and volatility, especially during banking crises \citep{jrfm17040134}.


Several studies explore whether Bitcoin can be used as a hedging tool against geopolitical risks, attempting to extrapolate risk premiums that could enhance investment portfolio returns \citep{BOURI2017192, BOURI2020156, LUTHER201750, Aysan2019, Long2022, bouri2022jumps, Mamum2020}.  \citet{Aysan2019} find a significant negative effect of changes in the GPR index on returns and a positive effect on Bitcoin's price volatility. They also document that GPR has a positive and statistically significant effect on Bitcoin's price volatility during bullish market periods. \citet{Long2022} find that coins with the lowest geopolitical beta outperform those with a high geopolitical beta. This result seems to be in line with those who suggest a certain safe-haven capacity for BTC and other safe-haven cryptocurrencies \citep{auer2022, jrfm17040134, GALATI2024103001}. In line with this hypothesis, \citet{bouri2022jumps} find that the price behavior of BTC jumps is dependent on jumps in the GPR index, supporting the idea that Bitcoin is a hedge against geopolitical risk. Conversely, \citet{Mamum2020} argue that BTC cannot be considered a safe-haven asset like gold, particularly during periods of significant economic and political uncertainty.

In developing countries, cryptocurrencies help address the lack of banking services, while in the United States and other Western countries, they are largely seen as speculative digital assets \citep{auer2022}. Supporting this notion, \citet{nouir2023economic} find that U.S. uncertainty impacts Bitcoin volatility in the short term, whereas Chinese uncertainty influences it in the long term. In particular, in developing countries, Bitcoin seems to offer viable alternatives to traditional financial systems. For example, in Turkey, 45\% of the population expects to own cryptocurrencies in the near future, compared to only 20\% in developed countries. Cryptocurrencies can serve as a response to the devaluation of national currencies and international sanctions \citep{ozdamar2021}.

While these studies provide valuable insights, they often overlook the differential impact of GPR on Bitcoin trading volumes across various countries and the nuanced distinctions between developing and developed economies. This gap highlights the need for a comprehensive analysis that considers these factors. The investigated hypothesis in this paper are twofold:
\begin{itemize}
    \item \textbf{H1:} Checking the impact of GPR on the growth rate of trading volumes at the country level;
    \item \textbf{H2:} Comparing the effects of GPR on BTC trading volume growth rates between developed and developing countries.
\end{itemize}

This paper is structured as follows: Section \ref{ch:Methodology} provides a comprehensive introduction to the dataset and the methodology employed, while Section \ref{ch:Analysis} illustrates our results. The discussion is presented in Section \ref{ch:Discussion}, and finally, Section \ref{ch:conclusions} offers concluding remarks.

\section{Data and Methodology}
\label{ch:Methodology}


This paper analyzes the monthly BTC trading volume, expressed in US dollars, in 33 countries and Europe, including (1) Argentina, (2) Australia, (3) Brazil, (4) Canada, (5) Switzerland, (6) Chile, (7) China, (8) Colombia, (9) Denmark, (10) Egypt, (11) Great Britain, (12) Hong Kong, (13) Hungary, (14) Indonesia, (15) India, (16) Japan, (17) Korea, (18) Mexico, (19) Malaysia, (20) Norway, (21) Peru, (22) Philippines, (23) Poland, (24) Russia, (25) Saudi Arabia, (26) Sweden, (27) Thailand, (28) Turkey, (29) Ukraine, (30) United States, (31) Europe, (32) Venezuela, (33) Vietnam, and (34) South Africa. This selection provides a global perspective on the crypto market and its internal dynamics, featuring diverse viewpoints from various countries and regions. The data is sourced from the \href{https://coin.dance/volume}{coin.dance} website. To our knowledge, no research has yet utilized this data, as our literature review found no such studies. Consequently, we believe this represents a novel approach to the subject.

\begin{figure}[h]
    \centering
    \includegraphics[width=0.9\linewidth]{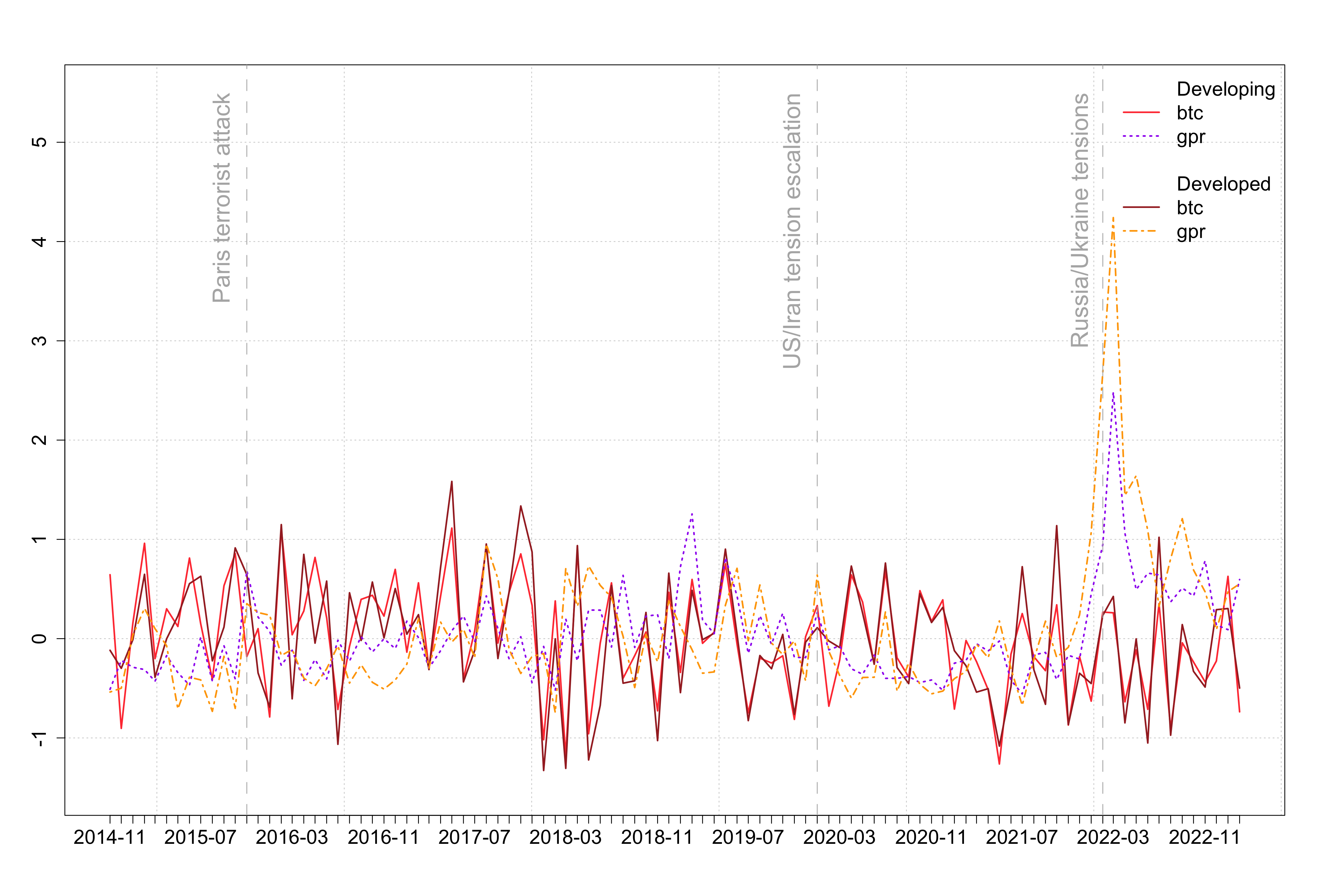}
    \caption{Standardized BTC trading volume growth ($btc$) and geopolitical risk index ($gpr$) over time for developing and developed countries. The gray dotted lines mark key events: the Paris terrorist attack, US/Iran tension escalation, and Russia/Ukraine tensions, respectively}
    \label{fig:volume_growth}
\end{figure}

Our sample includes monthly data from December 2014 to February 2023, resulting in an unbalanced panel dataset of 3271 observations. Through testing, we confirm that the variables employed  are stationary and that there is no multicollinearity among the explanatory variables.\footnote{See Fig.\ref{fig:volume_growth} for standardized BTC trading volume growth (BTC) and the geopolitical risk index (GPR) over time for developing and developed countries.}

While the use of monthly and national trading volume data limits our ability to include controls such as the VIX or OVX, incorporating country and monthly fixed effects strengthens our results (respectively \(\gamma_{i}\) and \(\theta_{t}\)) (Eq. \eqref{eq:panel}).
This approach allows us to account for all internal characteristics of different countries and control for temporal variations. For example, using  fixed effects we control for Gross Domestic product (GDP), unemployment rate, country regulations, and fixed situations across different countries, as well as external shocks like COVID-19 on a monthly basis. Our model is therefore as follows:
\begin{equation}
\label{eq:panel}
\textit{btc}_{i,t} = \beta_{0} + \beta_{1} \textit{btc}_{i,t-1} + \beta_{2} \textit{gpr}_{i,t} + \beta_{3} \textit{infl}_{i,t} + \beta_{4} \textit{change}_{i,t} + \beta_{5} \textit{share}_{i,t} + \gamma_{i} + \theta_{t} + \epsilon_{i,t}
\end{equation}
In Eq.\eqref{eq:panel}, the dependent variable represents the BTC trading volume growth rate by country. We include \( \textit{btc}_{i,t-1} \) to help control for autocorrelation, which is often present in datasets with long time series. This approach offers several benefits, including improving model accuracy, reducing potential biases, checking for and controlling autocorrelation, and effectively capturing the temporal dynamics inherent in time-series data.

Our aim is to examine the impact of the geopolitical risk index (GPR), denoted as \textit{gpr}$_{i,t}$, on \( \textit{btc}_{i,t-1} \). The GPR, constructed by \citet{Caldara22}, quantifies adverse geopolitical events and associated risks through an analysis of newspaper articles covering geopolitical tensions. The data for this index is sourced from the Economic Policy Uncertainty website. To enhance the quality and robustness of our estimates, we incorporate several control variables.


We collect data for our control variables from different sources (see Table \ref{tab:description}). We consider inflation (\(\textit{infl}_{i,t}\)) an important factor that can push  to diversify  wealth into BTC,  the monthly inflation growth rate  gathered from the International Monetary Fund (IMF). Another important indicator for our analysis is the growth rate of the country’s share index (\(\textit{share}_{i,t}\)). For this, we collect monthly data from the Organization for Economic co-operation and development (OECD).\footnote{Share price indices are calculated from the prices of common shares of companies traded on national or foreign stock exchanges.}  
Finally, we calculate the monthly change of the local currency with respect to BTC (\(\textit{change}_{i,t}\)). Although BTC is not considered an international currency like the dollar, we posit that it can serve as a valid indicator of the strength of national currencies. The transition to a standard measure, in this case BTC, is necessary to account for the inherent strength or weakness of the local currency.
In Table \ref{tab:descriptive}, we present some descriptive statistics of the variables employed in this analysis.\footnote{All growth rates are calculated as logarithmic differences.}




\begin{table}[ht]
\centering
\caption{Variable description.}
{\scriptsize
\begin{tabular}{llll}
\toprule
Variable          & Description                                                      & Unit     & Source         \\
\midrule
$\textit{btc}_{i,t}$               & BTC trading volume growth rate by country                                         &  \%       & From \emph{coin.dance}\footnote{\url{https://coin.dance/volume}} to author's elab..
    \\
$\textit{gpr}_{i,t}$   & Index measuring adverse geopolitical events and associated risks &  Index    & \cite{Caldara22} \\
$\textit{infl}_{i,t}$  & Inflation growth rate                                    &  \%       & IMF            \\
$\textit{change}_{i,t}$ & Change of local currency with respect to BTC            &  \%       & Author's elaboration    \\
$\textit{share}_{i,t}$  & Growth rate of the country’s share index                &  \%       & OECD           \\
$\textit{developing}_{i,t}$  & List of developing countries    &  Dummy       & IMF           \\       
\bottomrule
\end{tabular}
\label{tab:description}
}
\end{table}

\begin{table}[ht]
\centering
\caption{Summary statistics.}
\footnotesize
\begin{tabular}{lrrrrrrr}
\toprule

Variable & Obs & Mean & Std. dev. & Min & Max &  Skewness & Kurtosis\\

\midrule
$\textit{btc}_{i,t}$       & 3271 & 0.01 & 0.46 & -5.30 & 4.10 & 0.17 & 21.02 \\
$\textit{gpr}_{i,t}$   & 3271 & 0.28 & 0.60 & 0.00 & 8.80 & 5.10 & 44.73 \\
$\textit{infl}_{i,t}$  & 2910 & 0.37 & 0.82 & -3.39 & 14.03 & 4.81 & 56.46 \\
$\textit{change}_{i,t}$ & 3271 & -0.05 & 0.20 & -1.19 & 2.07 & -1.23 & 11.67 \\
$\textit{share}_{i,t}$   & 2258 & 0.00 & 0.05 & -0.33 & 0.20 & -1.58 & 12.94 \\
$\textit{developing}_{i,t}$   & 3271 & 0.64 & 0.48 & 0.00 & 1.00 & -0.58 & 1.34 \\
\bottomrule
\end{tabular}
\label{tab:descriptive}
\end{table}

\subsection{IV} \label{IV}

The fixed-effects specification in Eq. \eqref{eq:panel} might be subject to estimation bias due to endogenous input choice. 
For example, $\textit{btc}_{i,t}$ may be influenced by market demand, regulation changes, technological innovation, cryptocurrency adoption, Bitcoin price volatility, competition among cryptocurrencies, speculation and investor behavior, and macroeconomic and geopolitical events.

To evaluate the possibility of reverse causality running from $\textit{btc}_{i,t}$ to $\textit{gpr}_{i,t}$, we implement an instrumental variable (IV) strategy. This involves predicting in a first-stage regression the level of $\textit{gpr}_{i,t}$ by including some average lags of $\textit{gpr}_{i,t}$ and incorporating the same fixed effects as in Eq. \eqref{eq:panel}. Using lagged values of $\textit{gpr}_{i,t}$ as instruments helps mitigate the endogeneity problem by leveraging past values that are presumably not contemporaneously correlated with the error term in the main equation, thus providing a source of exogenous variation \citep{villasboas1999endogeneity}.\footnote{We experiment with different sets of average lags (x): (1) -40 and -24, (2) -36 and -20, and (3) -35 and -23. See Table \ref{tab: first stage}.}

Here is the first-stage regression equation:

\begin{equation}
\label{eq:zero stage}
gpr_{i,t} = \beta_{0} + \beta_{1} \textit{gpr}_{i,t-(x)} + \gamma_{i} + \theta_{t} + \epsilon_{i,t}
\end{equation}

We use various lags of \(\textit{gpr}_{i,t-(x)}\) to obtain the prediction \(\textit{gpr}^*_{i,t}\) from Eq. \eqref{eq:zero stage}. Consequently, this predicted value serves as an instrument for the actual \(\textit{gpr}_{i,t}\) in the main regression equation (Eq. \eqref{eq:panel}).

\section{Empirical Results}
\label{ch:Analysis}

The purpose of this analysis is to examine the impact of geopolitical risk (GPR) on the growth rate of Bitcoin trading volume by country, utilizing a fixed effects panel model. The suitability of the fixed effects model over the random effects model is validated through the Hausman test.

Table \ref{tab:regres}
 presents the estimation results. Column (1) shows the results for the entire sample, indicating that GPR is not significant. Columns (2) and (3) provide the panel estimates for developed and developing countries, respectively. For developing countries (column (3)), GPR is significant with a coefficient of 0.019. In contrast, for developed countries (column (2)), GPR is negative (-0.048) and not significant.

In these contexts, it is prudent to conduct the same analysis while controlling for variables that could potentially influence the BTC trading volume growth rate. It is well established that factors such as the inflation rate, local currency exchange rate, and the country's stock index growth affect the dependent variable. Including these control variables ensures a more robust analysis \citep{IMF2022}.

Columns (4) to (9) in Table \ref{tab:regres} report the results of our estimates with these additional controls. The GPR coefficient remains consistently positive and significant for developing countries across all estimates, while it remains negative and not significant for developed countries.

To further validate our findings, columns (10) and (11) in Table \ref{tab:regres} display the results of a panel model excluding the lagged dependent variable. The estimation results are consistent with the previous findings.\footnote{Table \ref{tab: dummyGPR} interacts the country dummy with $\textit{gpr}_{i,t}$. From this relationship, it emerges that countries categorized as developing by the IMF tend to have a positive coefficient. This further confirms the results that emerge from this study.}

\begin{figure}[htbp]
    \centering
    \begin{subfigure}[b]{0.48\textwidth}
        \centering
        \includegraphics[width=\textwidth]{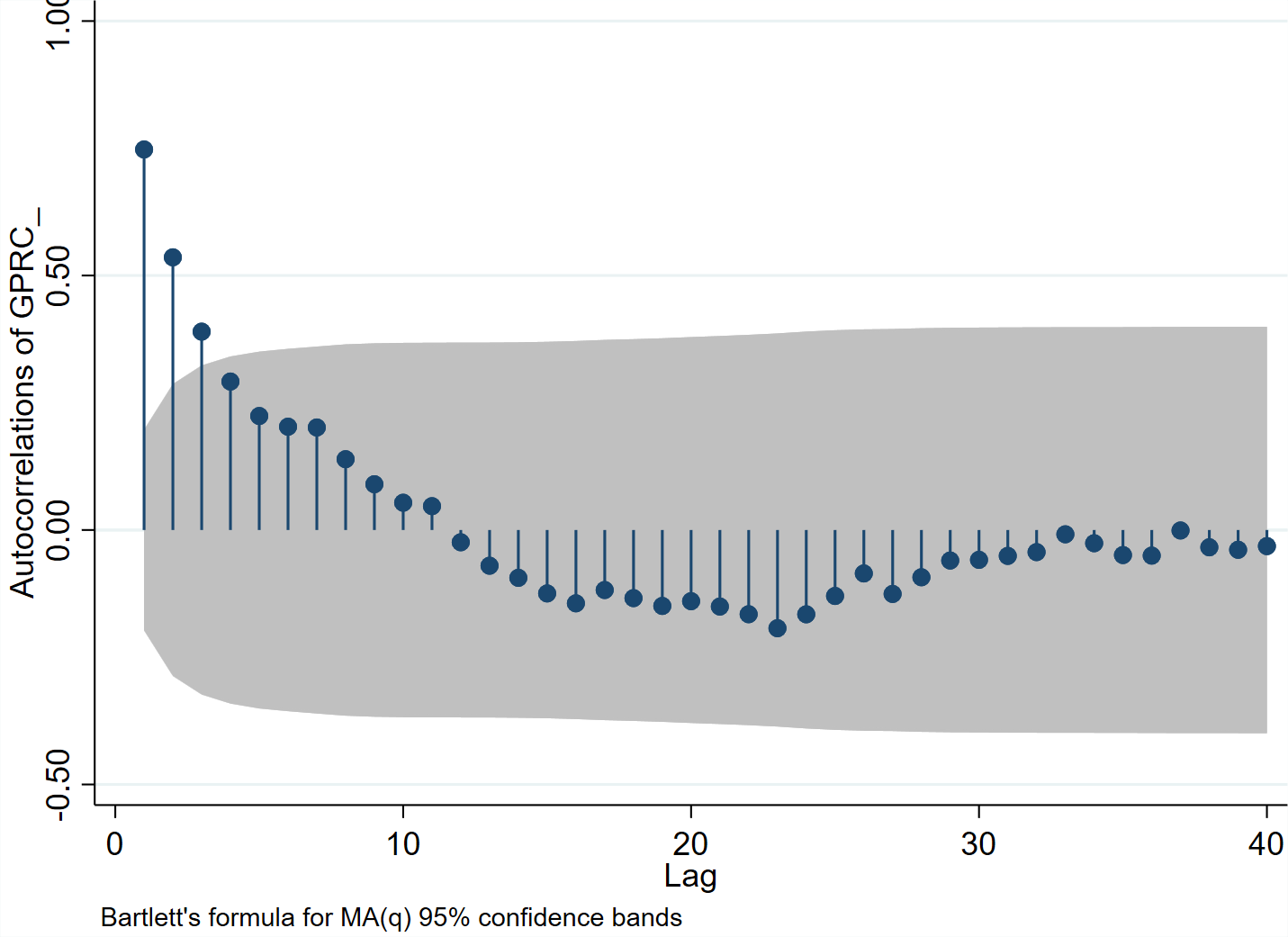}
        \caption{Autocorrelation}
        \label{fig:AC}
    \end{subfigure}
    \hfill
    \begin{subfigure}[b]{0.48\textwidth}
        \centering
        \includegraphics[width=\textwidth]{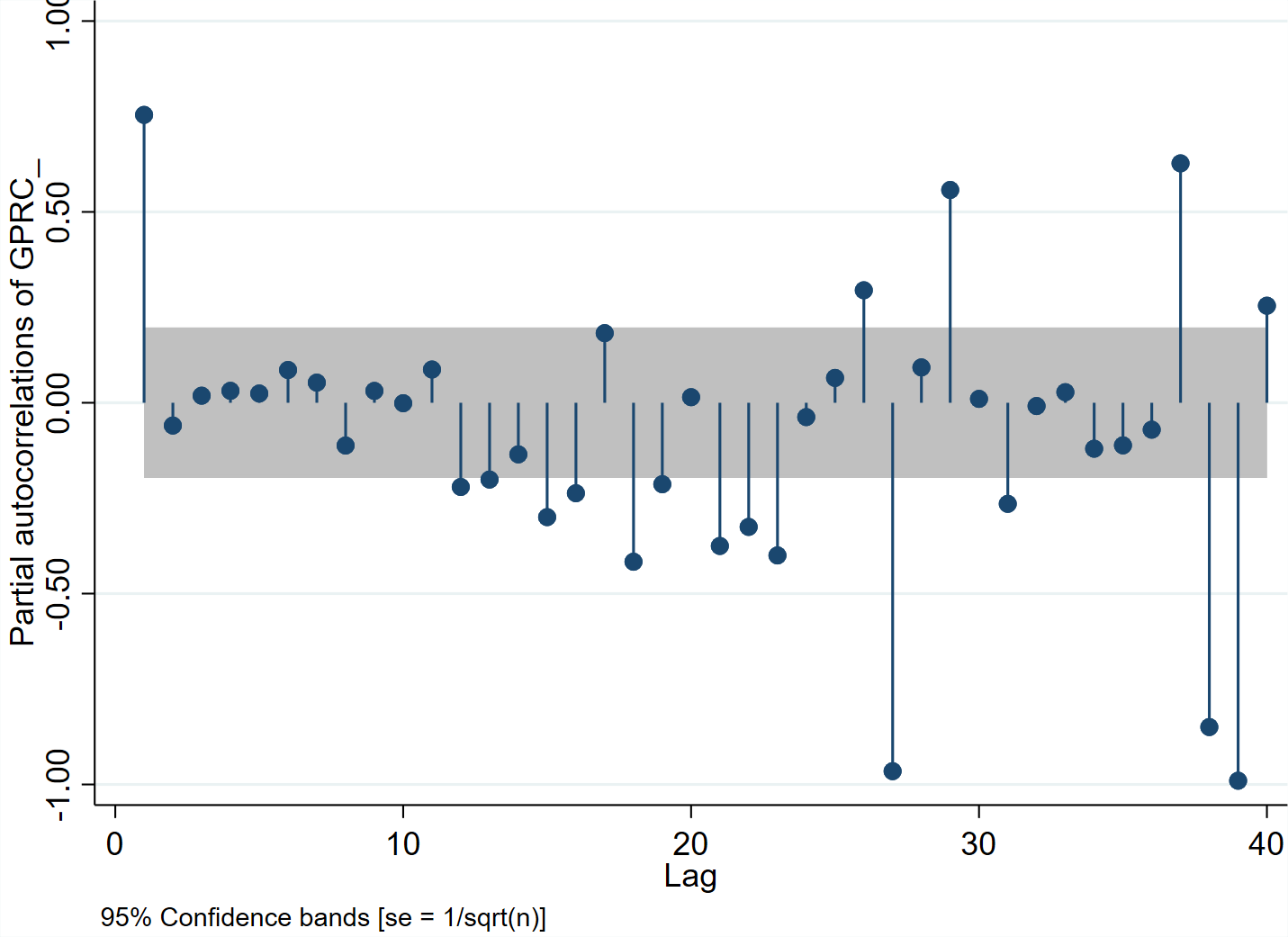}
        \caption{Partial Autocorrelation}
        \label{fig:PAC}
    \end{subfigure}
    \caption{Geopolitical Risk: Comparison of Autocorrelation and Partial Autocorrelation. The Confidence interval (CI) in (a) is based on Bartlett’s formula for MA(q) processes while the 95\% CI in (b) is calculated using a standard error of $1/\sqrt{n}$.}
    \label{fig:comparison}
\end{figure}

The preliminary results presented in Table \ref{tab:regres} are further validated by our robustness checks.
The instrumental variable (IV) analysis results in Table \ref{tab:IV}) enhances the robustness of our results, as discussed in Section \ref{IV}. Specifically, we observe that the coefficient increases by nearly three times compared to the reference values in column (9) of Table \ref{tab:regres}.
Additionally, Table \ref{tab:IV} shows a significant increase in the R-squared, indicating a better explanation of the phenomenon. The use of different lags further suggests the actual robustness of the results. These lags are chosen according to what emerges from the autocorrelation and partial correlation functions in Figure \ref{fig:comparison}. In terms of validity of our IVs, both the Underidentification Test (UIT) and the Weak Identification Test (WIT) provide important insights into the validity and strength of the instruments used. The UIT shows that the instruments are not underidentified, as all p-values  are below the 5\% significance level, confirming that the model is well-identified. The Cragg-Donald Wald F statistic (WIT) results indicate that the instruments are strong, with values consistently far exceeding the critical value of 6.66 for a 20\% maximal IV size. However, the Kleibergen-Paap rk Wald F statistic (see F statistic and CV 20\%) shows values that are close to or slightly below the 20\% critical value, suggesting that the instruments are generally strong.


 To ensure that our results are not biased due to autocorrelation, Table \ref{tab:regression_laggedbtc} presents the outcomes of the panel model specified in Eq. \eqref{eq:panel}, including various lags of the dependent variable, up to a maximum of four. The results remain consistent with those discussed in Table \ref{tab:regres}, and it is observed that the autocorrelation of $\textit{btc}_{i}$ diminishes after the fourth lag.

The Table \ref{tab: balanced panel} further verifies the results by implementing Eq. \eqref{eq:panel} using a balanced panel. Again, the results are confirmed. The variable $\textit{gpr}_{i,t}$ appears to be positively correlated with a higher growth rate of the volume of BTC. 

Another aspect we check is the control of volatility. By removing the months of extreme volatility (both upward and downward), we continue to enhance the quality of our estimates. High volatility may lead to spurious significance of the coefficients due to artificially inflated standard errors. The results so far are further confirmed by Table \ref{tab: perc}. In particular, comparing Column 9 in Table \ref{tab:regres} with Column 3 in Table \ref{tab: perc} reveals two key points: 1) when considering only the bulk of the data (80\%), we observe that the coefficients remain consistent with those from the full sample; 2) R-squared decreases by approximately 12\%, from 0.372 to 0.331. Therefore, the relationship between $\textit{gpr}_{i,t}$ and $\textit{btc}_{i,t}$ is not to be sought during periods of market euphoria or extreme fear, but rather in situations of normal market functioning.

As a further analysis (see the Appendix (\ref{SecA})), to determine if the impact of $\textit{gpr}_{i,t}$ on developing countries pertains only to BTC, we use $Y_{i,t}\equiv \textit{share}_{i,t}$ as the dependent variable , maintaining the same structure as Eq. \eqref{eq:panel}. The results are interesting: individuals from developing countries seem to be attempting to escape from their native currency while also avoiding their own firm's market share (see Table\ref{tab: share}).


\begin{sidewaystable} 
{\footnotesize
\centering
\caption{Impact of GPR on BTC trading volume growth rate by developed and developing countries. }
\label{tab:regres}
\begin{tabular}{lccccccccccc}
\toprule
 & (1) & (2) & (3) & (4) & (5) & (6) & (7) & (8) & (9) & (10) &(11) \\
VARIABLES & All sample & \cellcolor{gray!30}Developed & \cellcolor{gray!70}Developing & \cellcolor{gray!30}Developed & \cellcolor{gray!70}Developing & \cellcolor{gray!30}Developed & \cellcolor{gray!70}Developing & \cellcolor{gray!30}Developed & \cellcolor{gray!70}Developing & \cellcolor{gray!30}Developed & \cellcolor{gray!70}Developing \\
\midrule
$\textit{btc}_{i,t-1}$ & -0.309{***} & \cellcolor{gray!30}-0.392{***} & \cellcolor{gray!70}-0.232{***} & \cellcolor{gray!30}-0.392{**} & \cellcolor{gray!70}-0.233{***} & \cellcolor{gray!30}-0.382{***} & \cellcolor{gray!70}-0.229{***} & \cellcolor{gray!30}-0.384{**} & \cellcolor{gray!70}-0.297{***} & & \\
 & (0.074) & \cellcolor{gray!30}(0.121) & \cellcolor{gray!70}(0.053) & \cellcolor{gray!30}(0.121) & \cellcolor{gray!70}(0.054) & \cellcolor{gray!30}(0.111) & \cellcolor{gray!70}(0.055) & \cellcolor{gray!30}(0.120) & \cellcolor{gray!70}(0.048) & & \\
$\textit{gpr}_{i,t}$ & 0.001 & \cellcolor{gray!30}-0.048 & \cellcolor{gray!70}0.019{*} & \cellcolor{gray!30}-0.046 & \cellcolor{gray!70}0.038{***} & \cellcolor{gray!30}-0.049 & \cellcolor{gray!70}0.052{**} & \cellcolor{gray!30}-0.040 & \cellcolor{gray!70}0.085{**} & \cellcolor{gray!30}-0.069 & \cellcolor{gray!70}0.088{***} \\
 & (0.012) & \cellcolor{gray!30}(0.051) & \cellcolor{gray!70}(0.011) & \cellcolor{gray!30}(0.052) & \cellcolor{gray!70}(0.011) & \cellcolor{gray!30}(0.052) & \cellcolor{gray!70}(0.019) & \cellcolor{gray!30}(0.052) & \cellcolor{gray!70}(0.029) & \cellcolor{gray!30}(0.057) & \cellcolor{gray!70}(0.027) \\
$\textit{infl}_{i,t}$ & & & & \cellcolor{gray!30}-0.007 & \cellcolor{gray!70}-0.003 & \cellcolor{gray!30}-0.010 & \cellcolor{gray!70}0.005 & \cellcolor{gray!30}-0.005 & \cellcolor{gray!70}0.002 & \cellcolor{gray!30}-0.007 & \cellcolor{gray!70}0.003 \\
 & & & & \cellcolor{gray!30}(0.022) & \cellcolor{gray!70}(0.010) & \cellcolor{gray!30}(0.022) & \cellcolor{gray!70}(0.011) & \cellcolor{gray!30}(0.032) & \cellcolor{gray!70}(0.016) & \cellcolor{gray!30}(0.062) & \cellcolor{gray!70}(0.015) \\
$\textit{change}_{i,t}$ & & & & & & \cellcolor{gray!30}0.573 & \cellcolor{gray!70}1.089{*} & \cellcolor{gray!30}0.594 & \cellcolor{gray!70}1.154 & \cellcolor{gray!30}0.810 & \cellcolor{gray!70}1.201 \\
 & & & & & & \cellcolor{gray!30}(0.539) & \cellcolor{gray!70}(0.606) & \cellcolor{gray!30}(0.552) & \cellcolor{gray!70}(1.053) & \cellcolor{gray!30}(0.775) & \cellcolor{gray!70}(0.953) \\
$\textit{share}_{i,t}$ & & & & & & & & \cellcolor{gray!30}-0.074 & \cellcolor{gray!70}0.047 & \cellcolor{gray!30}0.346 & \cellcolor{gray!70}0.027 \\
 & & & & & & & & \cellcolor{gray!30}(0.796) & \cellcolor{gray!70}(0.207) & \cellcolor{gray!30}(1.134) & \cellcolor{gray!70}(0.206) \\
$(intercept)$  & 0.018{***} & \cellcolor{gray!30}0.021 & \cellcolor{gray!70}0.023{***} & \cellcolor{gray!30}0.026 & \cellcolor{gray!70}0.025{***} & \cellcolor{gray!30}0.053 & \cellcolor{gray!70}0.070{**} & \cellcolor{gray!30}0.051 & \cellcolor{gray!70}0.071 & \cellcolor{gray!30}0.078 & \cellcolor{gray!70}0.063 \\
 & (0.004) & \cellcolor{gray!30}(0.021) & \cellcolor{gray!70}(0.003) & \cellcolor{gray!30}(0.023) & \cellcolor{gray!70}(0.006) & \cellcolor{gray!30}(0.042) & \cellcolor{gray!70}(0.026) & \cellcolor{gray!30}(0.045) & \cellcolor{gray!70}(0.041) & \cellcolor{gray!30}(0.057) & \cellcolor{gray!70}(0.036) \\
\midrule
Observations & 3,271 & \cellcolor{gray!30}1,175 & \cellcolor{gray!70}2,096 & \cellcolor{gray!30}977 & \cellcolor{gray!70}1,933 & \cellcolor{gray!30}977 & \cellcolor{gray!70}1,933 & \cellcolor{gray!30}878 & \cellcolor{gray!70}1,171 & \cellcolor{gray!30}890 & \cellcolor{gray!70}1,183 \\
R-squared & 0.306 & \cellcolor{gray!30}0.383 & \cellcolor{gray!70}0.300 & \cellcolor{gray!30}0.385 & \cellcolor{gray!70}0.293 & \cellcolor{gray!30}0.398 & \cellcolor{gray!70}0.303 & \cellcolor{gray!30}0.397 & \cellcolor{gray!70}0.372 & \cellcolor{gray!30}0.259 & \cellcolor{gray!70}0.308 \\
Country FE & YES & YES & YES & YES & YES & YES & YES & YES & YES & YES & YES \\
Month FE & YES & YES & YES & YES & YES & YES & YES & YES & YES & YES & YES \\
\bottomrule
\end{tabular}

\textit{Robust standard errors in parentheses. *** p<0.01, ** p<0.05, * p<0.1}
}
\end{sidewaystable}
\begin{table}[htbp]
\centering
\caption{Robustness Check 1: Lagged IV Estimation Using Different Average Lags of GPR as instrument.}
\label{tab:IV}
{\footnotesize
\begin{tabular}{lcccccc}
\toprule
 & (1) & (2) & (3) & (4) & (5) & (6) \\
 VARIABLES& \cellcolor{gray!30}Developed & \cellcolor{gray!70}Developing & \cellcolor{gray!30}Developed &\cellcolor{gray!70} Developing & \cellcolor{gray!30}Developed & \cellcolor{gray!70}Developing \\
\midrule
$\textit{btc}_{i,t-1}$   & \cellcolor{gray!30}-0.377** &\cellcolor{gray!70} -0.305*** & \cellcolor{gray!30}-0.386** & \cellcolor{gray!70}-0.300*** & \cellcolor{gray!30}-0.385** & \cellcolor{gray!70} -0.308*** \\
                         & \cellcolor{gray!30}(0.127) & \cellcolor{gray!70}(0.045) & \cellcolor{gray!30}(0.126) & \cellcolor{gray!70}(0.049) &\cellcolor{gray!30} (0.126) & \cellcolor{gray!70}(0.046) \\
$\textit{gpr*}_{i,t}$    &\cellcolor{gray!30} 0.058 & \cellcolor{gray!70}0.294** & \cellcolor{gray!30}0.048 & \cellcolor{gray!70}0.251* & \cellcolor{gray!30} 0.133 &\cellcolor{gray!70}0.287** \\
                         &\cellcolor{gray!30} (0.102) &\cellcolor{gray!70} (0.131) & \cellcolor{gray!30}(0.078) & \cellcolor{gray!70}(0.119) & \cellcolor{gray!30}(0.103) & \cellcolor{gray!70}(0.095) \\
$\textit{infl}_{i,t}$    & \cellcolor{gray!30}-0.021 &\cellcolor{gray!70} 0.010 & \cellcolor{gray!30}-0.020 &\cellcolor{gray!70} 0.009 & \cellcolor{gray!30}-0.017 & \cellcolor{gray!70}0.010 \\
                         &\cellcolor{gray!30} (0.032) &\cellcolor{gray!70} (0.017) &\cellcolor{gray!30} (0.032) &\cellcolor{gray!70} (0.017) & \cellcolor{gray!30}(0.032) & \cellcolor{gray!70}(0.016) \\
$\textit{change}_{i,t}$  &\cellcolor{gray!30} 0.468 & \cellcolor{gray!70}1.243 &\cellcolor{gray!30} 0.517 &\cellcolor{gray!70} 1.201 & \cellcolor{gray!30}0.477 & \cellcolor{gray!70}1.225 \\
                         &\cellcolor{gray!30} (0.462) & \cellcolor{gray!70}(1.088) & \cellcolor{gray!30}(0.493) &\cellcolor{gray!70} (1.089) & \cellcolor{gray!30}(0.468) & \cellcolor{gray!70}(1.111) \\
$\textit{share}_{i,t}$   &\cellcolor{gray!30} -0.259 & \cellcolor{gray!70}-0.211 &\cellcolor{gray!30} 0.101 &\cellcolor{gray!70} -0.243 &\cellcolor{gray!30} 0.183 &\cellcolor{gray!70}-0.223 \\
                         &\cellcolor{gray!30} (0.833) & \cellcolor{gray!70}(0.268) &\cellcolor{gray!30} (0.646) & \cellcolor{gray!70}(0.238) & \cellcolor{gray!30}(0.670) & \cellcolor{gray!70}(0.279) \\
$(intercept)$            &\cellcolor{gray!30} -0.001 &\cellcolor{gray!70} 0.020 & \cellcolor{gray!30}0.005 & \cellcolor{gray!70}0.032 & \cellcolor{gray!30}-0.039 & \cellcolor{gray!70}0.024 \\
                         &\cellcolor{gray!30} (0.045) & \cellcolor{gray!70}(0.060) & \cellcolor{gray!30}(0.035) & \cellcolor{gray!70}(0.057) & \cellcolor{gray!30}(0.047) & \cellcolor{gray!70}(0.061) \\
\midrule
Observations & 841 & 1,105 & 873 & 1,149 & 849 & 1,116 \\
R-squared & 0.398 & 0.390 & 0.400 & 0.379 & 0.400 & 0.383 \\
Country FE & YES & YES & YES & YES & YES & YES \\
Month FE & YES & YES & YES & YES & YES & YES \\
UIT & 0.047 & 0.047 & 0.048 & 0.048 & 0.047 & 0.047 \\
WIT & 120.407 & 120.407 & 139.296 & 139.296 & 110.849 & 110.849 \\
F statistic & 8.661 & 8.661 & 6.912 & 6.912 & 7.534 & 7.534 \\
CV20\%  & 6.66 & 6.66 & 6.66 & 6.66 & 6.66 & 6.66 \\
Instrument & Lag -40 -24 & Lag -40 -24 & Lag -36 -20 & Lag -36 -20 & Lag -35 -23 & Lag -35 -23 \\
\bottomrule
\end{tabular}

\textit{Robust standard errors in parentheses. *** p<0.01, ** p<0.05, * p<0.1}
}
\end{table}

\begin{table}[ht]
\centering
\caption{Robusteness Check 2: Impact of lagged dependent variable on estimation results on developing countries. Fixed effect panel.}
{\footnotesize
\label{tab:regression_laggedbtc}
\begin{tabular}{lcccc}
\toprule
VARIABLES & (1) & (2) & (3) & (4) \\
\midrule
$\textit{btc}_{i,t-1}$          & -0.297{***} & -0.346{***} & -0.342{***} & -0.299{***} \\
             & (0.048)      & (0.064)      & (0.065)      & (0.059)      \\
$\textit{btc}_{i,t-2}$        &              & -0.179{***} & -0.229{***} & -0.247{***} \\
             &              & (0.052)      & (0.071)      & (0.077)      \\
$\textit{btc}_{i,t-3}$        &              &              & -0.149{***} & -0.148{***} \\
             &              &              & (0.045)      & (0.039)      \\
$\textit{btc}_{i,t-4}$        &              &              &              & -0.021       \\
             &              &              &              & (0.027)      \\
$\textit{gpr}_{i,t}$       & 0.085{**}   & 0.086{**}   & 0.078{***}  & 0.076{**}   \\
             & (0.029)      & (0.029)      & (0.024)      & (0.027)      \\
$\textit{infl}_{i,t}$       & 0.002        & -0.007       & -0.005       & -0.007       \\
             & (0.016)      & (0.013)      & (0.014)      & (0.013)      \\
$\textit{change}_{i,t}$  & 1.154        & 1.025        & 0.830        & 0.926        \\
             & (1.053)      & (0.994)      & (0.795)      & (0.901)      \\
$\textit{share}_{i,t}$    & 0.047        & 0.098        & 0.119        & 0.074        \\
             & (0.207)      & (0.260)      & (0.294)      & (0.254)      \\
$(intercept)$      & 0.071        & 0.075        & 0.075{*}    & 0.081{*}    \\
             & (0.041)      & (0.043)      & (0.038)      & (0.043)      \\
\midrule
Observations & 1,171        & 1,159        & 1,147        & 1,135        \\
R-squared    & 0.372        & 0.389        & 0.411        & 0.414        \\
Country FE   & YES          & YES          & YES          & YES          \\
Month FE     & YES          & YES          & YES          & YES          \\
\bottomrule
\end{tabular}

\textit{Robust standard errors in parentheses. *** p<0.01, ** p<0.05, * p<0.1}
}
\end{table}

\begin{table}[!htbp] 
\centering
\caption{Robusteness check 3: Balanced panel.}
{\footnotesize
\begin{tabular}{lccc}
\hline
 & \multicolumn{1}{c}{(1)} & \multicolumn{1}{c}{(2)} & \multicolumn{1}{c}{(3)} \\
VARIABLES & All sample & \cellcolor{gray!30}Developed & \cellcolor{gray!70}Developing \\ \hline
$\textit{btc}_{i,t-1}$ & -0.449*** & \cellcolor{gray!30}-0.482*** & \cellcolor{gray!70}-0.301** \\
 & (0.091) & \cellcolor{gray!30}(0.094) &\cellcolor{gray!70} (0.086) \\
$\textit{gpr}_{i,t}$ & 0.024 &\cellcolor{gray!30} 0.002 & \cellcolor{gray!70}0.295** \\
 & (0.029) & \cellcolor{gray!30}(0.038) &\cellcolor{gray!70} (0.097) \\
$\textit{infl}_{i,t}$ & -0.006 &\cellcolor{gray!30} -0.023 &\cellcolor{gray!70} 0.006 \\
 & (0.009) &\cellcolor{gray!30} (0.037) & \cellcolor{gray!70}(0.006) \\
$\textit{change}_{i,t}$ & 0.595 &\cellcolor{gray!30} 0.540 &\cellcolor{gray!70} 0.584* \\
 & (0.514) & \cellcolor{gray!30}(0.523) &\cellcolor{gray!70} (0.226) \\
$\textit{share}_{i,t}$ & -0.324 & \cellcolor{gray!30}-0.562 & \cellcolor{gray!70}0.073 \\
 & (0.445) & \cellcolor{gray!30}(0.745) & \cellcolor{gray!70}(0.224) \\
$(intercept)$  & 0.039 & \cellcolor{gray!30}0.030 & \cellcolor{gray!70}0.020* \\
 & (0.022) & \cellcolor{gray!30}(0.035) &\cellcolor{gray!70} (0.008) \\
\hline
Observations & 1,176 & 686 & 490 \\
R-squared & 0.417 & 0.448 & 0.522 \\
Country FE & YES & YES & YES \\
Month FE & YES & YES & YES \\
\hline
\end{tabular}
\label{tab: balanced panel}

\textit{Robust standard errors in parentheses. *** p<0.01, ** p<0.05, * p<0.1}
}
\end{table}

\begin{table}[htbp]
\centering
\caption{Robusteness check 4: Regression results for the 10th to 90th (Volatility Check) percentile sample of $\textit{btc(gr)}_{i,t}$. }
{\footnotesize
\label{tab: perc}
\begin{tabular}{lccc}
\toprule
 & (1) & (2) & (3) \\
 & All sample & \cellcolor{gray!30}Developed & \cellcolor{gray!70} Developing \\
\midrule
$\textit{btc}_{i,t-1}$ & -0.362*** &\cellcolor{gray!30} -0.437*** &\cellcolor{gray!70} -0.259*** \\
    & (0.083)   & \cellcolor{gray!30}(0.105)   & \cellcolor{gray!70}(0.041)   \\
$\textit{gpr}_{i,t}$  & 0.017   &\cellcolor{gray!30} -0.050   & \cellcolor{gray!70}0.098**  \\
      & (0.031)  & \cellcolor{gray!30}(0.054)  & \cellcolor{gray!70}(0.041)  \\
$\textit{infl}_{i,t}$ & -0.012  & \cellcolor{gray!30}-0.022  &\cellcolor{gray!70} -0.002   \\
      & (0.014)  & \cellcolor{gray!30}(0.033)  & \cellcolor{gray!70}(0.016)  \\
$\textit{change}_{i,t}$ & 0.725   &\cellcolor{gray!30} 0.542   &\cellcolor{gray!70} 1.474   \\
          & (0.605)  &\cellcolor{gray!30} (0.544)  & \cellcolor{gray!70}(1.267) \\
$\textit{share}_{i,t}$ & -0.017  &\cellcolor{gray!30} -0.006  & \cellcolor{gray!70}0.046   \\
         & (0.255)  & \cellcolor{gray!30}(0.939)  & \cellcolor{gray!70}(0.216) \\
$(intercept)$  & 0.105**  &\cellcolor{gray!30} 0.099   & \cellcolor{gray!70}0.149*  \\
         & (0.038)  &\cellcolor{gray!30} (0.056)  & \cellcolor{gray!70}(0.076) \\
\midrule
Observations & 1,821  & 780    & 1,041   \\
R-squared    & 0.313  & 0.388  & 0.331   \\
Country FE   & YES    & YES    & YES     \\
Month FE     & YES    & YES    & YES     \\
\bottomrule
\end{tabular}

\textit{Robust standard errors in parentheses. *** p<0.01, ** p<0.05, * p<0.1}
}
\end{table}

\clearpage

\section{Discussion} \label{ch:Discussion}

Our findings lead us to assert that geopolitical risk is associated with an increase in BTC volumes in developing countries. This suggests that in less advantaged nations, BTC volume is correlated with GPR. Specifically, when internal tensions rise, this trend becomes evident with an increase in volumes.

Therefore, while we reject Hypothesis 1, which posits whether geopolitical risk leads to increased trading volumes, we cannot reject Hypothesis 2: in developing countries, BTC is incrementally exchanged during periods of high geopolitical risk.

This result is very important as it provides a clearer picture of the perception of BTC around the world. While in developed countries BTC remains a speculative asset with limited real-world implications, its perception can drastically change in other parts of the world, potentially serving as a different option compared to the traditional exchange markets. In developing countries, fears of national currency devaluation, high inflation, or internal and international economic shocks drive people towards seeking an alternative, even if only temporary, to the local currency.

To the best of our knowledge, this is the first study to analyze BTC volumes at the country level, making the available literature for comparison limited. However, \citet{ZHANG2023103620} finds a significant and positive correlation between geopolitical risk (GPR) and the volatility of emerging stock markets, indicating a heightened sensitivity to geopolitical risk in developing countries. This study supports our findings. Given the relationship between volatility and uncertainty, the increase in BTC volumes appears to be an unusual reaction to market fears.




\clearpage
\section{Conclusions}
\label{ch:conclusions}

This paper explores the impact of geopolitical risk (GPR) on BTC volume growth rates. Our sample considers 33 individual countries and the European economic region and uses the news-based GPR index from \citet{Caldara22}. Our findings indicate that GPR has a significant positive effect on the growth rate of BTC volumes in developing countries, whereas this relationship does not appear to exist in the wealthiest regions. Additionally, GPR significantly and positively affects BTC volume when inflation, exchange rates, and stock price shares are included as control variables, thus confirming the robustness of our results.

Given that trading volume reflects interest in an asset, it is evident that interest in the primary cryptocurrency (BTC) is growing in developing countries. Typically, an increase in volume is followed by a rise in price \citep{balcilar2017can}, indicating a higher level of buying activity compared to selling activity.

These findings may have significant implications for policymakers and market investors. First of all, the GPR is an important factor in explaining Bitcoin (BTC) volume growth rates. Policymakers should consider the GPR if they aim to regulate the cryptocurrency market, especially in developing countries. The second point raises a new question rather than providing an answer: Why do emerging countries increase their interest in Bitcoin when the GPR rises? Understanding the reasons behind this trend can be particularly important at the country level. We believe there could be multiple answers to this question. For example, people may want to protect their wealth in a different currency/good due to inflation, war, corruption, religious considerations, or strict banking controls.

We suggest that future research could explore the spatial effects of Bitcoin trading volume across different regions. While Bitcoin can act as a hedge against various risks, people might also choose to invest in more stable currencies like USD or Euro, or in commodities such as gold and silver. For this reason, spatial effects could be important: the proximity between different countries, exchange rates, ease of conversion, and remittances. Analyzing these spatial patterns can deepen our understanding of the increasing global interest in cryptocurrencies.



\clearpage

\clearpage
\appendix
\section{Appendix}\label{SecA}

\begin{table}[ht]
\centering
\caption{Fixed Effects Regression Results.}
{
\begin{tabular}{l c c l}
\toprule
VARIABLES & Coefficient & Standard Error & Classification \\
\midrule
y = L & -0.337*** & (0.071) & \\
\rowcolor{gray!70} ARG\#$\textit{gpr}_{i,t}$ & 0.407** & (0.157) & Developing \\
\rowcolor{gray!30} AUS\#$\textit{gpr}_{i,t}$ & 0.264*** & (0.074) & Developed \\
\rowcolor{gray!70} BRA\#$\textit{gpr}_{i,t}$ & 0.177 & (0.223) & Developing \\
\rowcolor{gray!30} CAN\#$\textit{gpr}_{i,t}$ & -0.468*** & (0.052) & Developed \\
\rowcolor{gray!30} CHE\#$\textit{gpr}_{i,t}$ & -1.131*** & (0.195) & Developed \\
\rowcolor{gray!70} CHL\#$\textit{gpr}_{i,t}$ & 2.207*** & (0.398) & Developing \\
\rowcolor{gray!70} CHN\#$\textit{gpr}_{i,t}$ & -0.027 & (0.016) & Developing \\
\rowcolor{gray!70} COL\#$\textit{gpr}_{i,t}$ & -0.030 & (0.226) & Developing \\
\rowcolor{gray!30} DNK\#$\textit{gpr}_{i,t}$ & 0.205 & (0.259) & Developed \\
\rowcolor{gray!70} EGY\#$\textit{gpr}_{i,t}$ & 0.044 & (0.223) & Developing \\
\rowcolor{gray!30} EUR\#$\textit{gpr}_{i,t}$ & -0.022 & (0.053) & Developed \\
\rowcolor{gray!30} GBR\#$\textit{gpr}_{i,t}$ & -0.034** & (0.013) & Developed \\
\rowcolor{gray!70} HKG\#$\textit{gpr}_{i,t}$ & 0.106* & (0.053) & Developing \\
\rowcolor{gray!70} HUN\#$\textit{gpr}_{i,t}$ & 0.541*** & (0.090) & Developing \\
\rowcolor{gray!70} IDN\#$\textit{gpr}_{i,t}$ & 0.660*** & (0.147) & Developing \\
\rowcolor{gray!70} IND\#$\textit{gpr}_{i,t}$ & 0.130* & (0.068) & Developing \\
\rowcolor{gray!30} JPN\#$\textit{gpr}_{i,t}$ & -0.581*** & (0.054) & Developed \\
\rowcolor{gray!30} KOR\#$\textit{gpr}_{i,t}$ & -0.276*** & (0.022) & Developed \\
\rowcolor{gray!70} MEX\#$\textit{gpr}_{i,t}$ & -0.070 & (0.125) & Developing \\
\rowcolor{gray!70} MYS\#$\textit{gpr}_{i,t}$ & -2.321*** & (0.259) & Developing \\
\rowcolor{gray!30} NOR\#$\textit{gpr}_{i,t}$ & 0.986*** & (0.095) & Developed \\
\rowcolor{gray!70} PER\#$\textit{gpr}_{i,t}$ & 0.072 & (0.295) & Developing \\
\rowcolor{gray!70} PHL\#$\textit{gpr}_{i,t}$ & 0.793*** & (0.210) & Developing \\
\rowcolor{gray!70} POL\#$\textit{gpr}_{i,t}$ & 0.174*** & (0.023) & Developing \\
\rowcolor{gray!70} RUS\#$\textit{gpr}_{i,t}$ & -0.004 & (0.005) & Developing \\
\rowcolor{gray!70} SAU\#$\textit{gpr}_{i,t}$ & 0.038 & (0.052) & Developing \\
\rowcolor{gray!30} SWE\#$\textit{gpr}_{i,t}$ & 0.250*** & (0.073) & Developed \\
\rowcolor{gray!70} THA\#$\textit{gpr}_{i,t}$ & -0.892*** & (0.240) & Developing \\
\rowcolor{gray!70} TUR\#$\textit{gpr}_{i,t}$ & 0.448*** & (0.029) & Developing \\
\rowcolor{gray!70} UKR\#$\textit{gpr}_{i,t}$ & 0.019*** & (0.004) & Developing \\
\rowcolor{gray!30} USA\#$\textit{gpr}_{i,t}$ & 0.010 & (0.008) & Developed \\
\rowcolor{gray!70} VEN\#$\textit{gpr}_{i,t}$ & -0.077 & (0.088) & Developing \\
\rowcolor{gray!70} VNM\#$\textit{gpr}_{i,t}$ & -0.346 & (0.208) & Developing \\
\rowcolor{gray!70} ZAF\#$\textit{gpr}_{i,t}$ & 0.607** & (0.242) & Developing \\
(Intercept) & 0.020*** & (0.004) & \\
\midrule
Observations & & 3,233 & \\
R-squared & & 0.321 & \\
Month FE & & YES & \\
\bottomrule
\end{tabular}

\begin{minipage}{\textwidth}
\centering
\textit{Robust standard errors in parentheses. *** p<0.01, ** p<0.05, * p<0.1}
\end{minipage}
}
\label{tab: dummyGPR}
\end{table}

\begin{table}[ht]
\centering
\footnotesize
\caption{Classification of Developed and Developing Countries According to the World Economic Outlook, IMF. Source: \href{https://www.imf.org/en/Publications/WEO/weo-database/2023/April/groups-and-aggregates}{IMF classification}.}
\begin{tabular}{llcll}
\toprule
\multicolumn{2}{c}{Developed} & & \multicolumn{2}{c}{Developing} \\
\cmidrule(r){1-2} \cmidrule(l){4-5}
Alpha-3  Code & Country & & Alpha-3  Code & Country \\
\cmidrule(r){1-2} \cmidrule(l){4-5}
AUS & Australia & & ARG & Argentina \\
CAN & Canada & & BRA & Brazil \\
CHE & Switzerland & & CHL & Chile \\
DNK & Denmark & & CHN & China \\
EUR & Euro Area & & COL & Colombia \\
GBR & United Kingdom & & EGY & Egypt \\
JPN & Japan & & HUN & Hungary \\
KOR & Korea & & IDN & Indonesia \\
NOR & Norway & & IND & India \\
SWE & Sweden & & MEX & Mexico \\
USA & United States & & MYS & Malaysia \\
HKG & Hong Kong & & PER & Peru \\
    &  & & PHL & Philippines \\
    &  & & POL & Poland \\
    &  & & RUS & Russia \\
    &  & & SAU & Saudi Arabia \\
    &  & & THA & Thailand \\
    &  & & TUR & Turkey \\
    &  & & UKR & Ukraine \\
    &  & & VEN & Venezuela \\
    &  & & VNM & Vietnam \\
    &  & & ZAF & South Africa \\
\bottomrule
\end{tabular}
\label{tab:country_distribution}
\end{table}


    \cellcolor{gray!70}
\begin{table}[htbp]
\centering
\footnotesize
\caption{Regression Results using share as dependent variable.}
\begin{tabular}{lccc}
\hline
                    & (1)       & (2)       & (3)       \\
           & All Sample        & \cellcolor{gray!30}Developed        & \cellcolor{gray!70}Developing        \\
\hline
$\textit{share}_{i,t-1}$         & 0.252***  &\cellcolor{gray!30} 0.210***  & \cellcolor{gray!70}0.237***  \\
                    & (0.038)   & \cellcolor{gray!30}(0.024)   & \cellcolor{gray!70}(0.040)   \\
$\textit{gpr}_{i,t}$               & -0.016**  & \cellcolor{gray!30}-0.001    & \cellcolor{gray!70}-0.026*** \\
                    & (0.007)   & \cellcolor{gray!30}(0.003)   & \cellcolor{gray!70}(0.002)   \\
$\textit{infl}_{i,t}$              & 0.001     &\cellcolor{gray!30} -0.001    & \cellcolor{gray!70}0.003     \\
                    & (0.002)   & (\cellcolor{gray!30}0.002)   &\cellcolor{gray!70} (0.002)   \\
$\textit{change}_{i,t}$         & 0.010     & \cellcolor{gray!30}-0.006    & \cellcolor{gray!70}0.079*    \\
                    & (0.009)   &\cellcolor{gray!30} (0.007)   & \cellcolor{gray!70}(0.040)   \\
$\textit{btc}_{i,t}$ & -0.000    & \cellcolor{gray!30}0.000     &\cellcolor{gray!70} 0.000     \\
                    & (0.001)   & \cellcolor{gray!30}(0.002)   & \cellcolor{gray!70}(0.002)   \\
$(intercept)$          & 0.010***  & \cellcolor{gray!30}0.004**   & \cellcolor{gray!70}0.013***  \\
                    & (0.003)   & \cellcolor{gray!30}(0.002)   & \cellcolor{gray!70}(0.003)   \\
\hline
Observations        & 2,069     & 887       & 1,182     \\
R-squared           & 0.588     & 0.788     & 0.564     \\
Country FE          & YES       & YES       & YES       \\
Month FE            & YES       & YES       & YES       \\
\hline
\end{tabular}

\textit{Robust standard errors in parentheses. *** p<0.01, ** p<0.05, * p<0.1}
\label{tab: share}
\end{table}

\begin{table}[h!]
    \centering
    \footnotesize
     \caption{First stage: dependent variable $\textit{gpr}_{i,t}$.}
    \begin{tabular}{lccc}
        \toprule
        VARIABLES & (1) & (2)  & (3) \\
         & All Sample        & All Sample          & All Sample          \\
        \midrule
       $\textit{gpr}_{i,t(-24;-40)}$ & -0.761*** & & \\
        & (0.257) & & \\
       $\textit{gpr}_{i,t(-20;-36)}$ & & -0.775** & \\
        & & (0.295) & \\
        $\textit{gpr}_{i,t(-23;-35)}$ & & & -0.681*** \\
        & & & (0.247) \\
        $(intercept)$  & 0.470*** & 0.470*** & 0.450*** \\
        & (0.062) & (0.070) & (0.059) \\
        \midrule
        Observations & 3,082 & 3,211 & 3,112 \\
        R-squared & 0.699 & 0.705 & 0.700 \\
        Country FE & YES & YES & YES \\
        Month FE & YES & YES & YES \\
        \bottomrule
    \end{tabular}

    \textit{Robust standard errors in parentheses. *** p<0.01, ** p<0.05, * p<0.1}
    \label{tab: first stage}
\end{table}

\end{document}